\documentstyle[prl,aps,twocolumn]{revtex}
\begin{document}
\draft
{\noindent \large \bf Comment on ``New Class of Resonances at the
Edge of the Two-Dimensional Electron Gas.''}
\vskip .5cm 
Recently, Zhitenev {\it et al}.\cite{Zhitenev} reported measurements
 of the capacitance of a gate covering the edge and part of a 2DEG in
 the quantized Hall regime. They observed a reproducible resonance
 structure as a function of magnetic field $B$ and gate voltage $V_g$.
 This structure was attributed to the variation in conductance of an
 incompressible strip separating the states of the $N$th and $N+1$th
 Landau levels (LL). Although the authors disfavored the explanation of
 the conductance resonances based on resonant tunneling (RT) through
 an impurity state, I believe that it should not be dismissed and can
 explain a number of experimental observations very well.  In
 particular, I point out a new characteristic signature of RT in the
 quantized Hall regime: the quantization of the slopes of the
 resonance tracks on the $B-V_g$ plot.

The authors argued that provided each resonance is due to a different
impurity along the edge the similarity in structures for different
sample lengths is puzzling. Contrary to their assumptions I suggest
that a {\it single} location along the edge is responsible for each
series of peaks in the resonance structure resolving the puzzle.  In
this location the long-range disorder potential creates in the middle
of the incompressible strip a droplet of electrons (holes) belonging
to the $N+1$th ($N$th) LL.

 This situation is similar to the conventional Coulomb blockade (CB)
with the incompressible region playing the role of an insulator
surrounding the charged island.  The existence of CB in the presence
of extended lower LL states has been convincingly demonstrated by
Alphenaar {\it et al.}\cite{Alphenaar}.  As the gate voltage is varied
the number of charges in the droplet changes giving rise to a series
of resonance peaks. The spacing between these peaks is determined by
the CB condition,
\begin{equation}
\label{cap}
\Delta V_g=e/C,
\end{equation}
yielding a periodic structure when the number of charges in the
droplet is large.  By using Eq.(\ref{cap}) the size of the droplet can
be estimated. For $\Delta V_g\approx 40$mV taken from
Ref.\cite{Zhitenev} I find the droplet to be about $600\times 600$
\AA. A droplet of this size may contain from one to ten electrons
(holes). A relatively small number of particles leads to fluctuations
in capacitance which may account for the observed deviations from
periodic peak structure\cite{Goldman}.

A major strength of the RT picture is its ability to explain the peak
tracks in the $B-V_g$ plot.  The CB resonance condition requires
having a half-integer number of particles in the droplet. As the
magnetic field is varied, the density of electrons on the completely
filled LL in the incompressible strip changes proportionally. To keep
the number of particles in the droplet fixed, the total electron
density in the droplet vicinity, $n$, must be adjusted:
\begin{equation}
\label{tracks}
n=NeB/hc+n_0.
\end{equation}
Here $n_0$ gives the average density of electrons (holes) on the
$N+1$th ($N$th) LL in the droplet vicinity. It must be independent of
$V_g$ or $B$ for a given resonance and vary from resonance to
resonance. As the distance between the 2DEG and the gate is larger
than the typical screening radius, the gate voltage, $V_g$, is
proportional to the total density, $n$, over the geometrical
capacitance per unit area, $C_g$.  By using this I find from
Eq.(\ref{tracks})
\begin{equation}
\label{tracks1}
dV_g/dB=Ne/C_ghc.
\end{equation}
This result shows that the slopes of the peak tracks must be
quantized, with $N$ being the filling factor of the incompressible
strip.  Indeed, based on the available data from the authors of
Ref.\cite{Zhitenev} I find that the ratios of the slopes for $N=1,2,4$
in a given $V_g$ interval agree with Eq.(\ref{tracks1}) with an order
of $10\% $ accuracy.

 Zhitenev {\it et al.}\cite{Zhitenev} found that the resonance peak
conductances grow with temperature, in contradiction with the RT model
assuming that the tunneling amplitudes are temperature
independent. However there are at least two reasons to believe that
this assumption is unjustified. First, the thickness of the tunneling
barrier is determined by the width of the incompressible strip which
decreases with temperature.  Second, electron correlations lead to the
formation of fractional channels\cite{Chklovskii}, tunneling into
which has a non-trivial temperature dependence\cite{Wen}.

It is straightforward to extend the RT model to the fractional
quantized Hall regime. There one should expect resonance tracks to be
straight lines with quantized slopes given by the fractional filling
factor in the incompressible strip. Since the charge of the droplet
can now change by a fraction, Eq.(\ref{cap}) predicts a smaller
distance between resonance peaks.

I thank B.I. Halperin, L.S. Levitov, K.A. Matveev, and D.H. Cobden for
useful discussions, the authors of Ref.\cite{Zhitenev} for sharing
their results prior to publication, and the Aspen Center for
Physics. I was supported by the Harvard SOF and by NSF Grant DMR
94-16910.

\vskip .5cm 
\leftline{Dmitri B. Chklovskii} 
\leftline{\ \ \ Physics Department} 
\leftline{\ \ \ Harvard University}
\leftline{\ \ \ Cmbridge MA 02138}
\vskip .5cm 
\leftline{Received 30 August 1996}
\leftline{PACS numbers: 73.40.Hm, 73.40.Gk, 73.23.H}


\begin{references}
\vspace {-1.5cm}
\bibitem{Zhitenev} N. B. Zhitenev {\it et al}., Phys. Rev. Lett. {\bf
77}, 1833 (1996).
\bibitem{Alphenaar} B. W. Alphenaar {\it et al}., Phys. Rev. B {\bf
46}, 7236 (1992).
\bibitem{Goldman} Bo Su, and V. J. Goldman, Science {\bf 255},
 313 (1992).
\bibitem{Chklovskii} D. B. Chklovskii, Phys. Rev. B {\bf 51}, 9895
(1995).
\bibitem{Wen} X.-G. Wen, Int. J. Mod. Phys. B {\bf 6}, 1711 (1992).
\end{references}
\end{document}